\documentclass[10pt]{article}

\begin{document}

\renewcommand{\theequation}{\arabic{section}.\arabic{equation}}
\baselineskip=20pt
\setlength\arraycolsep{1pt}
\newfont{\largebx}{cmssbx10 scaled\magstep4}   

\title{\largebx Kerr cat states from the  four-photon Jaynes-Cummings model}

\author{ {Hongchen Fu\thanks{email: h.fu@open.ac.uk}  \ 
                    and Allan I Solomon\thanks{email: a.i.solomon@open.ac.uk}}\\
{\normalsize Quantum Processes Group, The Open University,  
                         Milton Keynes, MK7 6AA, UK}}
\maketitle

\begin{abstract}
We investigate the dynamics of  a four-photon Jaynes-Cummings 
model for large photon number. It is shown that at certain times
the cavity field is in a  pure state which is a superposition of two 
Kerr states, analogous to the Schr\"{o}dinger cat state (superposition 
of two coherent  states)  which occurs in the one and two photon 
cases. 
\end{abstract}

\section{Introduction}

The Jaynes-Cummings model (JCM) is a standard and  important 
model which describes  the interaction of an atom and 
a single mode radiation field in a cavity \cite{jay,sho}. The importance of
this model is due  not only to its exact solvability, but also arises from some of its
purely  quantum effects, such as the periodic
collapse and revival of the atomic number population,  and the behaviour of the 
cavity field which becomes a  {\em Schr\"{o}dinger cat} state - a 
phase-correlated superposition of two coherent states -  at 
certain times. The JCM model has been generalized 
in many different ways. One such is  the multiphoton 
generalization \cite{sho}
described by the following Hamiltonian ($\hbar =1$)
\begin{equation}
H=\omega a^{\dagger }a+\frac{1}{2}\omega _{0}\sigma _{3}+
g(a^{\dagger k}\sigma_{-} + a^{k} \sigma_{+}),
\end{equation}
where $k$ is a positive integer, 
$ \sigma_3=|e\rangle\langle e|-|g\rangle\langle g|$,
$ \sigma_+=|e\rangle\langle g| $ and $ \sigma_-=|g\rangle\langle e|$
are atomic operators, $ g $ is the coupling constant, $\omega
_{0}$ and $\omega $ are the atomic transition frequency and cavity resonant
mode frequency respectively. It is interesting that for both the one 
\cite{onep} and two \cite{buzek,fufs} photon cases, the cavity 
field is in a Schr\"{o}dinger cat state at intervals corresponding 
to one half of the revival time. This arises from the fact that  the 
Rabi frequency in both cases is proportional to $n$ (up to a constant) 
in the approximation of large photon number
\cite{onep,buzek,fufs}. 

In this paper we shall address  the dynamics and 
quantum characteristics of the 4-photon JCM in the large photon
number approximation. Under this approximation the 
generalized Rabi frequency (Eq.(2.3)) depends on $n$ nonlinearly
(up to $n^2$) and this nonlinearity causes Kerr nonlinearity
in the cavity. We shall study the photon number distribution
(Sec. 3) and the entropy of the cavity field (Sec.4) and the Q-function
(Sec. 5). It is found that the cavity field is in a Kerr state
or a macroscopic superposition of two Kerr states at certain
times. We also study the atomic number inversion  and find that it
exhibits collapse and revival for short time intervals.

\section{Large photon number approximation}
\setcounter{equation}{0}

We assume that at the initial time $t=0$, the atom and field are
decoupled and the atom is initially prepared in the excited state $|e\rangle $, while
the field is in the coherent state $|\alpha\rangle $
\begin{equation}
    |\alpha\rangle = \sum_{n=0}^\infty C_n |n\rangle, \qquad
    C_n = e^{-|\alpha|^2/2}\frac{\alpha^n}{\sqrt{n!}},
\end{equation}    
where $\alpha$ is a complex number.
For simplicity, we only consider the on-resonance interaction case
$\omega_0 = k \omega$ as in \cite{buzek}. Then the combined 
atom-field wave function 
at time $t$ is obtained as
\begin{equation} \label{wavefunc}
   |\Psi(\tau)\rangle = \sum_{n=0}^\infty C_n \left( \cos(\Omega_n \tau)
   |n,e\rangle - i \sin(\Omega_n \tau)|n+k,g\rangle \right),
\end{equation}   
where $\tau\equiv gt$ is the {\em scaled time} and $\Omega_n$ is
the {\em generalized Rabi frequency} \cite{rabi} (in units of $\omega_0$)
\begin{equation}
   \Omega_n \equiv \sqrt{(n+1)(n+2)\cdots (n+k)}.
\end{equation}   
In the  large photon number case, namely when 
the average photon number $\bar{n}$ is large enough 
($\bar{n} \gg k)$, the Poisson distribution $\{C_n\}$ is mainly 
concentrated near $\bar{n}$ and we have $n\sim \bar{n}$. 

We now specialize to the case $k=4$.
To facilitate an analytical treatment, we write the generalized
Rabi frequency $\Omega_n$ as
\begin{equation}
    \Omega_n = n^2 \sqrt{1-x}, \qquad
    x\equiv \frac{10}{n}+\frac{35}{n^2}
    +\frac{50}{n^3}+\frac{24}{n^4}.
\end{equation}    
For the large photon number case, $x$ is a small term and
we can use the following Taylor expansion
\begin{equation}
    \sqrt{1-x}=1+\frac{x}{2}-\frac{x^2}{8}+\cdots.
    \label{sqx}
\end{equation}
Here we can only neglect the terms in $\Omega_n$
which are much smaller than 1, namely, we must keep 
the terms to $n^{-2}$ in (\ref{sqx}). This is because
$\Omega_n$ is the argument of {\em trigonometric} functions 
in the wave function (\ref{wavefunc}) and a small constant 
(for example $\pi/2$) can 
effect the physical quantity drastically. We therefore  need 
to calculate the first three terms in (\ref{sqx}). We then 
 obtain the approximate generalized Rabi frequency 
\begin{equation}
    \Omega_n\approx n^2 +5n+5,  \qquad
    \Omega_{n-4} \approx n^2 -3n+1,
    \label{4app}
\end{equation}    
and $\Omega_n-\Omega_{n-4}=4(2n+1)$. Here
the constant terms (5 or 1) are much smaller than
$n^2$ but cannot be neglected. 

\section{Photon number distribution}
\setcounter{equation}{0}

Taking the trace in the atom space, one finds for
the reduced density operator of the cavity field 
\begin{equation} \label{pndexact}
   \rho_F=\sum_{m,n=0}^\infty C_n C_m^* \left[
   \cos(\Omega_n t)\cos(\Omega_m t)|n\rangle\langle m|+
   \sin(\Omega_n t)\sin(\Omega_m t)|n+4\rangle\langle m+4|
   \right],
\end{equation}   
from which the photon number distribution (PND) 
$P_n(\tau)\equiv \langle n|\rho_F|n\rangle $ is obtained as
\begin{eqnarray}
   P_n(\tau)&=&|C_n|^2 \cos^2(\Omega_n \tau) + 
   |C_{n-4}|^2 \sin^2(\Omega_{n-4}\tau)        \nonumber \\
   &\approx &|C_n|^2 \cos^2\left[\Omega_{n-4}\tau + 
   4(2n+1)\tau \right] + 
   |C_{n-4}|^2 \sin^2\left[\Omega_{n-4}\tau\right].
\end{eqnarray}
It is easy to see that the PND is a $\pi$-periodic function
of the scaled time $\tau$. This means that at $\tau=\pi$, we have 
$P_n(\pi)=|C_n|^2$, a Poisson distribution. At $\tau=\pi/2$, we have
$P_n(\pi/2)=|C_{n-4}|^2$, where we have used that fact that
$n^2-3n+1$ is always an odd integer. This means that the 
PND is also a Poisson distribution, but displaced by 4.
However, the wave function (pure state) is not necessarily a 
coherent state;  we shall see in the next section
that the wave function is actually a Kerr state \cite{Kerr}.

When $\tau=\pi/4$, we have $4(2n+1)\pi/4=(2n+1)\pi$ and
therefore
\begin{equation}
   P_n(t)=|C_n|^2 \cos^2\left[ (n^2 - 3n+1)\pi/4\right]
   +|C_{n-4}|^2  \sin^2\left[ (n^2 - 3n+1)\pi/4\right].
   \label{pi4}
\end{equation}
Writing $n=4m+i, \ i=0,1,2,3$, we have 
\begin{equation}
   (n^2-3n+1)\frac{\pi}{4}=(4m^2+2im-3m)\pi+
   \left\{ \begin{array}{ll}
             \frac{\pi}{4}  & \mbox{ for } i=0,3;\\
             -\frac{\pi}{4}  & \mbox{ for }  i=1,2.
            \end{array}
   \right.             
\end{equation}   
Then the PND is finally obtained as
\begin{equation}
    P_n(t)=\frac{1}{2}\left(|C_n|^2 +|C_{n-4}|^2\right),
\end{equation}   
namely, the average of PND at $\tau=0$ and $\pi/2$,
as shown in Fig.1.

When $\tau=\pi/8$, $\Omega_n-\Omega_{n-4}=
n\pi+\pi/2$ and the PND is obtained as 
\begin{equation}
   P_n(t)=\left(|C_n|^2 + |C_{n-4}|^2 \right)
   \sin^2\left[(n^2 -3n+1)\frac{\pi}{8}\right] ,
   \label{pi8}
\end{equation}
which is a strongly oscillating function; in other words,
the photon number distribution exhibits strong oscillation.
Writing $n=8p+i, \ i=0,1,\cdots,7$, we can further write
(\ref{pi8}) as 
\begin{equation}   
   P_n(t)= \left\{ \begin{array}{ll}
            \frac{\displaystyle 2-\sqrt{2}}{\displaystyle 4} 
            \left(|C_n|^2 + |C_{n-4}|^2 \right), &
                     \mbox{ for } i=0,1,2,3; \\
            \frac{\displaystyle 2+\sqrt{2}}{\displaystyle 4} 
            \left(|C_n|^2 + |C_{n-4}|^2 \right), &
                     \mbox{ for } i=4,5,6,7.
            \end{array}        \right. 
\end{equation}
So for any $n$ the photon number distribution is non-vanishing 
and, in that sense, the oscillation is not {\em perfect}. 
However, as can be seen  from Fig.1, 
at the slightly earlier time $\tau=\pi/8 - \xi$, the photon number
distribution exhibits perfect oscillation in that  it becomes zero  
at $n=8p$ and $8p+3$. 

\section{Entropy and pure states of the cavity field}

The entropy $S$ of a quantum-mechanical system 
is a measure of how close  the system is to a {\em pure}  
state and is defined 
by \cite{alfred,barnett}
\[
S=-\mbox{Tr} \left[\rho \ln (\rho )\right],
\]
where $\rho $ is the density operator of the quantum system
and the Boltzmann constant is assumed to be unity.
$S=0$ for a pure state and $S>0$ for a mixed state.
In our model, the initial state is prepared
in a pure state, so the whole atom-field system remains in a 
pure state at any time $t>0$ and its entropy is always zero. 
However, due to the 
entanglement of the atom and the cavity field at $t>0$, both the
atom and the field are generally in mixed states, although
at certain times the field and the atomic subsystems
are `almost' in pure states. 

Since the initial state is a pure state, the entropy $S_{F}$ 
of the cavity field equals  the atomic entropy $S_{A}$ 
\cite{barnett}. The entropy $S_F$ or $S_A$, which is
referred to as the {\em entanglement} of the total system in quantum
information, is used to measure the amount of entanglement 
between  the two subsystems. When
$S_F = S_A =0$, the system is {\em disentangled} or 
{\em separable} and both the field and atomic subsystems
are in pure states.

It is more convenient to calculate the entropy of the 2-level
atomic system. From Eq.(\ref{wavefunc}) 
the atomic reduced
density operator $\rho_{A}$ can be readily obtained as
\begin{equation}
     \rho _{A}(\tau)\equiv 
     \mbox{Tr}_{F}(|\Psi(\tau)\rangle\langle\Psi(\tau)|)
     =\rho _{11}|g\rangle
     \langle g|+\rho _{12}|g\rangle \langle e|+\rho_{21}
     |e\rangle \langle g|+\rho _{22}|e\rangle \langle e|,
\end{equation}
where
\begin{eqnarray}
     && \rho _{11}(\tau)=\sum_{n=0}^{\infty}|C_{n} |^{2}
                            \sin ^{2}(\Omega _{n}t) , \qquad
           \rho _{22}(\tau) =\sum_{n=0}^{\infty}|C_{n} |^{2}
                            \cos ^{2}(\Omega _{n}t),  \nonumber \\
     &&\rho _{12}(\tau) =\rho _{21}^{\ast }=\sum_{n=0}^{\infty}
         C_{n+4} C_{n} 
         \cos (\Omega _{n+4}t)\sin (\Omega_{n}t).  
\end{eqnarray}
The field and atomic entropy
$S_{A}=-\mbox{Tr}_{A}(\rho _{a}\ln (\rho _{a})) $
can be expressed as
\begin{equation}
     S_{F}=S_{A}=-\pi _{+}\ln (\pi _{+})-\pi _{-}\ln (\pi _{-}),
\end{equation}
where $\pi _{\pm}$ are the eigenvalues of the atomic reduced field
density operator $\rho_{A}$
\begin{equation}
    \pi_{\pm }=\frac{1}{2}(1\pm \sqrt{(\rho _{22}-
    \rho_{11})^{2}+4| \rho_{12}|^{2}}).
\end{equation}

Noting that $\Omega_n$ is an odd integer for large photon 
number, it is easy to see that  the entropy 
$S_{F}$ of the cavity field is a $\pi$-periodic function of $\tau$.
We  plot  the entropy $S_{F}$ as a function of 
$\tau$ for $\bar{n}=50$ and $0\leq \tau\leq \pi$ in Fig.2.  
We observe that the entropy is dynamically 
reduced to zero at $\pi/2$ and the cavity field can 
be periodically found in pure states. We also observe that the 
entropy falls quickly to minima {\em near} $ \tau=\pi/4, 3\pi/4$
(see also Fig.3).

Those calculated features can be explained analytically. The periodicity
of $S_F$ implies that $S_F(m\pi)=0$ for any positive integer $m$. 
In this case the wave function (\ref{wavefunc}) can be readily
obtained as
\begin{equation}
    \Psi(m\pi)\approx e^{im\pi} |\alpha\rangle \otimes |e\rangle;
\end{equation}
namely, the cavity field is in the coherent state $|\alpha\rangle$
(up to a phase).

When $\tau=\pi/2$, we have $\rho_{11}=1, 
\rho_{22}=\rho_{12}=0$ and thus the entropy $S_F(\pi/2)=0$.
In this case the wave function of the system is
\begin{equation}
    \Psi(\pi/2)=-i\left(\sum_{n=0}^\infty C_n (-1)^{n(n+1)/2}
    |n\rangle\right)\otimes |g\rangle
    = e^{-\pi i/2} |-\alpha,\pi\rangle\otimes|g\rangle;    
\end{equation}
that is,  the atom is in its ground state and the field is in
a pure state, which is known as a Kerr state \cite{Kerr}
\begin{equation}
    |\alpha, \gamma\rangle\equiv
    e^{-|\alpha|^2/2}\sum_{n=0}^\infty \frac{\alpha^n}{\sqrt{n!}}
    e^{i\gamma n(n-1)/2}|n\rangle.
\end{equation}    
When the atom is detected in its ground state, the cavity field 
will be in the Kerr state. 

When $\tau=\pi/4$ and $3\pi/4$, we have $\rho_{11}=\rho_{22}$ and
$\rho_{12}\approx 0$, where we have used the following
fact for large photon number
\begin{equation}
     \cosh(\bar{n}) = \sum_{n=0}^\infty \frac{\bar{n}^{2n}}{(2n)!}       
     \approx \sum_{n=0}^\infty \frac{\bar{n}^{2n+1}}{(2n+1)!}  =\sinh(\bar{n}).
\end{equation}                      
The entropy is then $S_F= -\ln(2)\approx 0.67315$ (The exact
numerical result is 0.69314, see Fig.3). This means that the cavity field
is in a mixed state. However, at a slightly earlier or later
time $\tau=\pi/4+\delta_r$, the entropy is dramatically 
reduced to a minimum, as  can be seen in  Fig.2 and Fig.3. For the specific choice
\begin{equation}
     \delta_r = \frac{r \pi}{16\bar{n}} \equiv r\delta_1, \qquad
     r=\pm1,\pm3,\cdots
\end{equation}
we obtain 
\begin{equation} \label{omegad}
     (\Omega_{n}-\Omega_{n-4})\left(\frac{\pi}{4}+\delta_r\right)
     \approx  (2n+r+1)\pi+\pi/2,
\end{equation}          
where we have used the relation $n \sim \bar{n}$ for large photon
number $\bar{n}$. In this case we can see that the entropy is 
approximately zero.
In fact, at $\tau=\pi/4+\delta_r$, using (\ref{omegad}) we obtain the 
following factorized state (writing $\alpha=|\alpha| e^{i\phi}$)
\begin{eqnarray}
\lefteqn{    |\Psi(\pi/4+\delta_r)\rangle \approx
    -\left(\sum_{n=0}^\infty C_n \sin(\Omega_n \tau)|n\rangle\right)
    \otimes \left( e^{4i\phi}|e\rangle + i|g\rangle\right)} \nonumber\\
    &=&\frac{i}{2}\left[ e^{i5({\pi\over 4}+\delta_r)}\left|
    -i\alpha e^{-i6\delta_r}, 
    \pi/2+2\delta_r\right\rangle -
    e^{-i5({\pi\over 4}+\delta_r)}\left|i \alpha e^{-i6\delta_r},
    -\pi/2-2\delta_r \right\rangle\right] \nonumber \\
    && \otimes
    \left( e^{4i\phi}|e\rangle + i|g\rangle\right),
\end{eqnarray}  
from which we see that the cavity field is a macroscopic
superposition of two Kerr states
\begin{equation}
    \frac{1}{\sqrt{2}}\left[ e^{i5({\pi\over 4}+\delta_r)}\left|
    -i\alpha e^{-i6\delta_r}, 
    \pi/2+2\delta_r\right\rangle -
    e^{-i5({\pi\over 4}+\delta_r)}\left|i \alpha e^{-i6\delta_r},
    -\pi/2-2\delta_r\right\rangle\right).
\end{equation}

The photon number distribution can be  derived from the
above wave function. Here we write down the PND directly
from (\ref{pndexact}) and (\ref{omegad})
\begin{equation}  \label{pnddelta1}
    P_n(\pi/4+\delta_r)=\left(|C_n|^2+|C_{n-4}|^2\right)
    \sin^2\left[(n^2-3n+1)(\pi/4+\delta_r)\right].
\end{equation}    
In Fig.4 we  compare  this
approximate result with the numerical result. We see that
the agreement is good.

We now  show that the entropy is approximately 
vanishing. Using $|C_n|^2 \approx |C_{n-4}|^2$ in
(\ref{pnddelta1}) we find
\begin{equation}
    1=\sum_{n=4}^\infty P_n(\pi/4+\delta_r)=
    2\sum_{n=0}^\infty |C_n|^2
    \sin^2\left[(n^2+5n+5)(\pi/4+\delta_r)\right]
    \equiv 2\rho_{11},
\end{equation} 
from which we have 
$\rho_{11}(\pi/4+\delta_r)\equiv1-\rho_{22}(\pi/4+\delta_r)=1/2$.   
>From (\ref{omegad}) we have
\begin{equation}
    \rho_{12}(\pi/4+\delta_r)= -e^{4i\phi}\rho_{11}(\pi/4+\delta_r)
    = -e^{4i\phi}/2.
\end{equation}    
So the entropy $S_F(\pi/4+\delta_r)\approx0$, as expected.

\section{The Q-function and Population Inversion}
\setcounter{equation}{0}

The quasi-probability distribution Q-function is defined as \cite{qfunc}:
\begin{equation} \label{qfuncdef}
     Q(\beta )=\frac{1}{\pi }\langle \beta |\rho_F |\beta \rangle,
\end{equation}
where $\rho_F $ is the reduced density operator of the 
cavity field given in
Eq.\,(\ref{pndexact}) and $|\beta\rangle$ is a coherent state.
Inserting Eq.\,(\ref{pndexact}) into Eq.\,(\ref{qfuncdef}) we can
easily obtain the Q-function of the cavity field
\begin{equation}
   Q(\beta)=\frac{e^{-|\beta|^2}}{\pi}\left(\left|
   \sum_{n=0}^\infty\frac{\beta^{*n}}{\sqrt{n!}}C_n
   \cos(\Omega_n \tau)\right|^2 +
   \left|
   \sum_{n=0}^\infty\frac{\beta^{*{n+4}}}{\sqrt{(n+4)!}}C_n
   \cos(\Omega_n \tau)\right|^2 \right).
\end{equation} 

When $\tau=0$ and $\pi$, the cavity field is in a coherent
state and its Q-function is Poissonian, as  can be seen from
Fig.\,5. At $\tau=\pi/2$, the Q-function has two components.
At $\tau=\pi/8$, the Q-function has 8 components, and the
field is in a mixed state (its entropy is 0.6888). So the 
interference between components results in {\em imperfect} 
oscillation of the PND.

When $\tau=\pi/4$ the Q-function is composed of 4 well-separated
components, the entropy reaches its maximum and the field
is in a mixed state. In a short  time $\tau=\pi/4+\delta_1$, the
field is nearly pure and the Q-function separates into 8 
components, the interference between which resulting in strong oscillation
of the PND, as  can be seen from Fig.(4). However, at 
$\tau=\pi/4+\delta_3$ and $\tau=\pi/4+\delta_5$
the cavity field is less pure and the Q-function has less well-separated
components, and the oscillation of the PND is less marked
(see Fig.4).

Finally, we briefly describe the atomic population inversion (API), given by
\begin{equation}
   W(\tau)=\langle\sigma_3\rangle=
   \sum_{n=0}^\infty |C_n|^2 \cos(2\Omega_n \tau).
\end{equation} 
In the one or two photon cases this exhibits the usual periodic
collapse and revival phenomenon. The 4-photon system also
shows this behaviour.  In Fig.6 we give a plot of
the API $\langle \sigma_3 \rangle $ for $\bar{n}=50$,
from which one can see that the API for the 4-photon 
system also exhibits  periodic collapse and revival.
However, due to the nonlinearity of the system, this 
effect does not last as long as in the one or two photon
cases.

\section{Conclusions}
\setcounter{equation}{0}

In this paper we  studied the four-photon JC model and
its dynamical behaviour. In the large photon number 
approximation, the Rabi frequency $\Omega_n$ depends 
quadratically on $n$  and this leads
to  nonlinear effects in the cavity.

By examining the entropy of the field or atom we found 
that the system is approximately disentangled at certain 
times. In this case both the atom and cavity field are
in a pure state and those pure states are explicitly given.
It is interesting that the cavity field exists  in both a nonlinear 
Kerr state and a macroscopic superposition of Kerr
states, analogous to the   coherent state and the Schr\"{o}dinger
cat states in the one or two photon JC model. One may 
refer to this analogous Kerr superposition as a Kerr cat state.

We analysed the PND in the large photon number approximation. 
At certain times the PND exhibits
strong oscillation. From the Q-function we know that 
the cavity field at those times has eight components; it is 
the interference between those components which causes
 the strong oscillation of the PND.  Note that these oscillations depend more sensitively on the interaction time than in the one or two photon cases due to the quadratic  dependence of the generalized Rabi frequency on photon number  in the present work. 

We also considered atomic number inversion and 
found that the phenomenon of periodic collapse and revival 
occurs; it is however a short-lived phenomenon  due to the 
effects of nonlinearity.   

Experimental implementation of the theoretical idea 
herein proposed would possibly depend on the  
use of a trapped neutral atom or ion in a high-Q 
cavity, or an atomic beam in transit through the 
cavity \cite{newrefs}. In any realistic scheme 
off-resonant effects must be taken into account, 
and for this reason the atomic species used must 
provide a good approximation to a two-level system.

\section*{Acknowledgements}

We are grateful to Dr.\,Andrew Greentree for  useful 
discussion and comment, especially with regard to the 
possible experimental realisation of the scheme proposed 
here. H.\,Fu is supported in part 
by the National Natural Science Foundation of China. 
A.\,I.\,S.  thanks the Laboratoire de Physique Th\'{e}orique 
des Liquides, Paris University VI, for  hospitality.


\newpage

\input epsf
\begin{figure} \label{fig-PND}
\begin{center}
\epsfbox{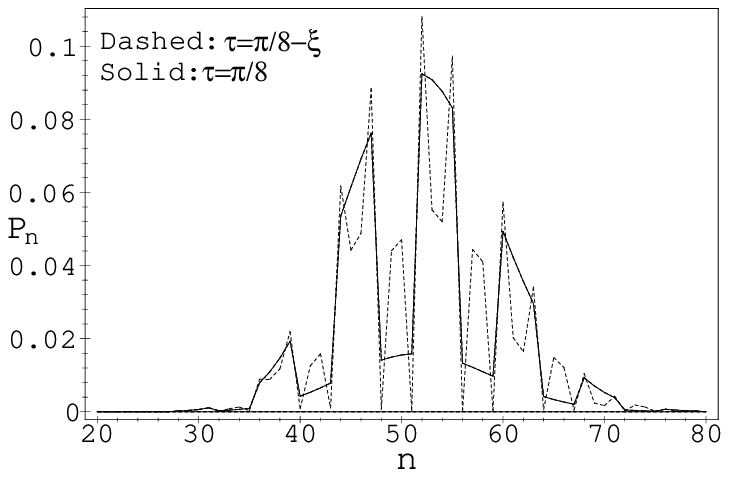}
\epsfbox{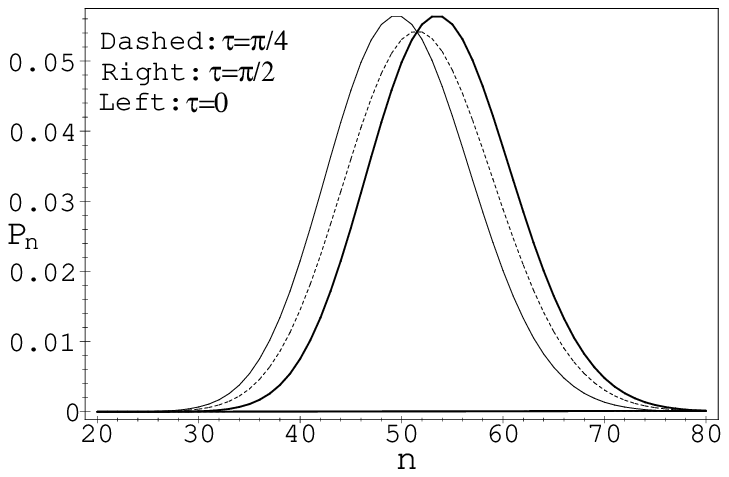}
\caption{Photon number distributions for $\bar{n}=50$
at $\tau=0, \pi/8, \pi/4$ and $\pi/24$. For the $\tau=\pi/8$ case, the
photon number distribution at slightly earlier time 
$\tau=\pi/8-\pi/24000$ (dashed line) is also presented.}
\end{center}
\end{figure}

\begin{figure}
\begin{center}
\epsfbox{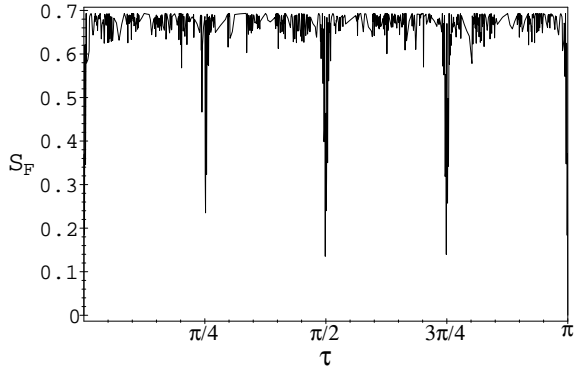}
\caption{Entropy $S_F(\tau)$ of the cavity field against the
scaled time $\tau$ for $\bar{n}=50$.}
\end{center}
\end{figure}

\begin{figure} 
\begin{center}
\epsfbox{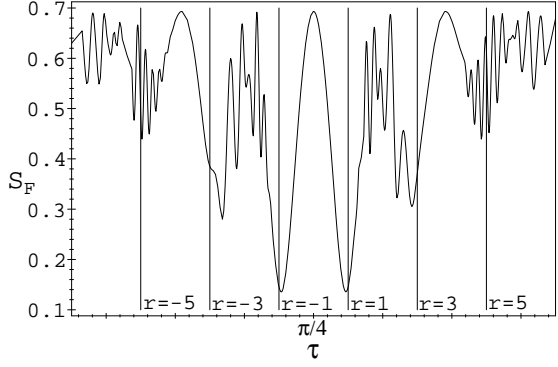}
\caption{Entropy $S_F(\tau)$ of the cavity field around 
$\tau=\pi/4$ for $\bar{n}=50$. The vertical lines with
labels $r=\pm 1, \pm 3, \pm 5$ represent $\tau=\pi/4+
r\delta_1$, $\delta_1=\pi/800$. }
\end{center}
\end{figure}

\begin{figure} \label{pnddfig}
\begin{center}
\epsfbox{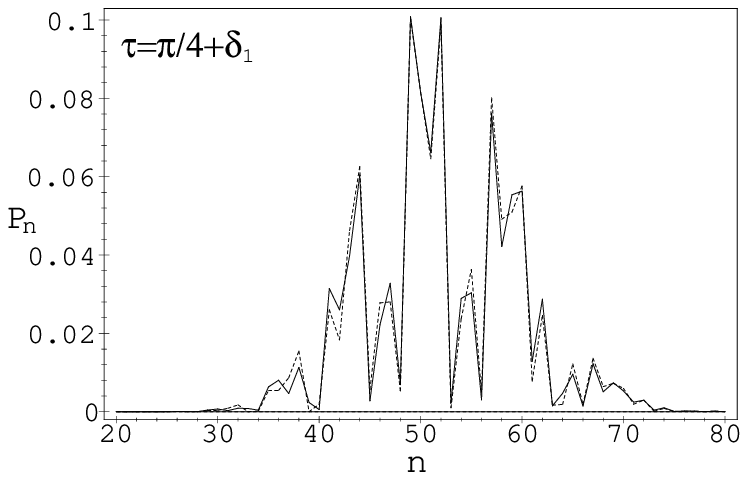}
\epsfbox{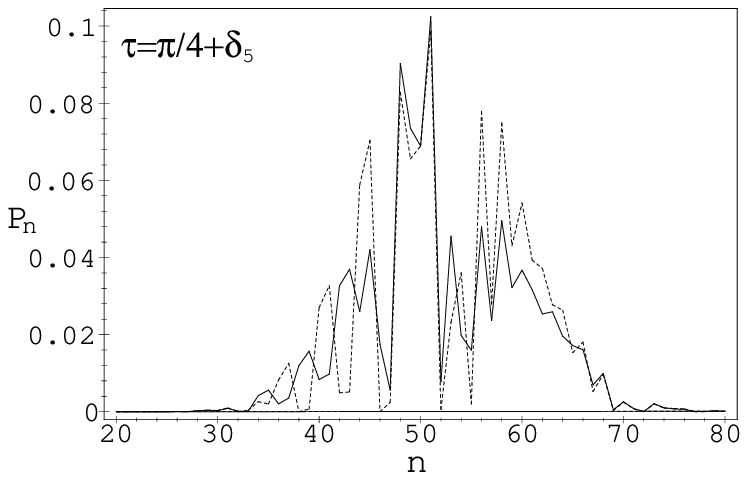}
\caption{Photon number distribution of the cavity field at 
$\tau=\pi/4+\delta_1$ and $\tau=\pi/4+\delta_5$ for $\bar{n}=50$. 
The solid line is the 
exact numerical result and the dashed line is from the 
approximate result Eq.(\ref{pnddelta1}).}
\end{center}
\end{figure}

\newpage

\begin{figure} \label{qfunc}
\begin{center} 
\centerline{ 
\epsfxsize=6cm
\epsfbox{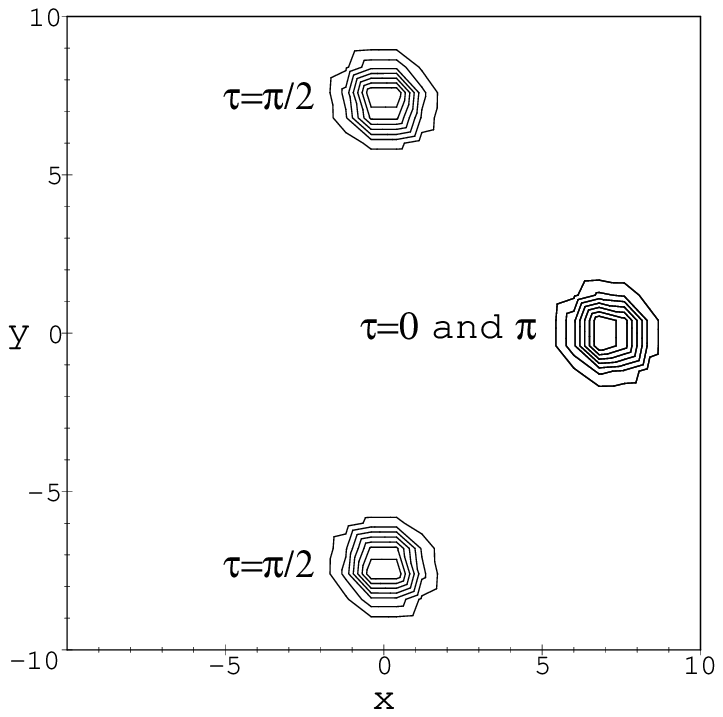}
\epsfxsize=6cm
\epsfbox{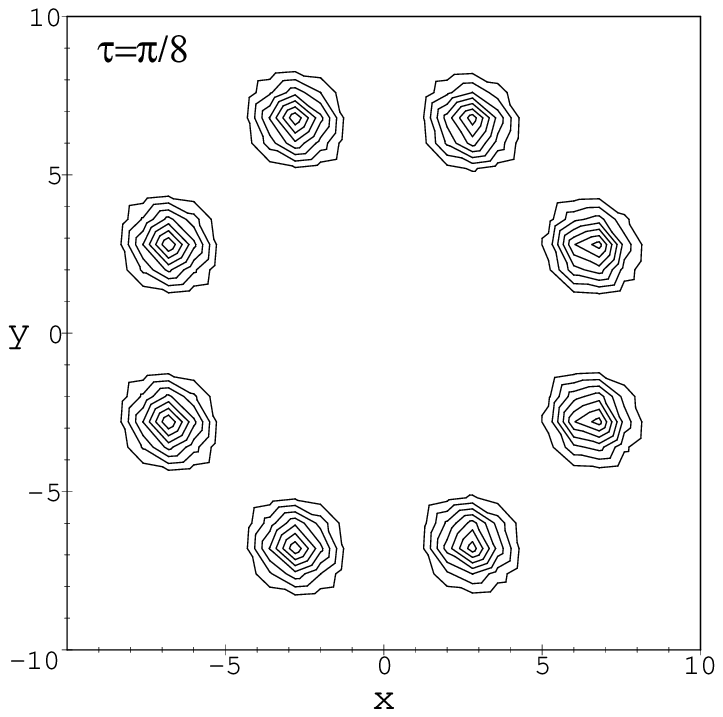}}
\centerline{
\epsfxsize=6cm
\epsfbox{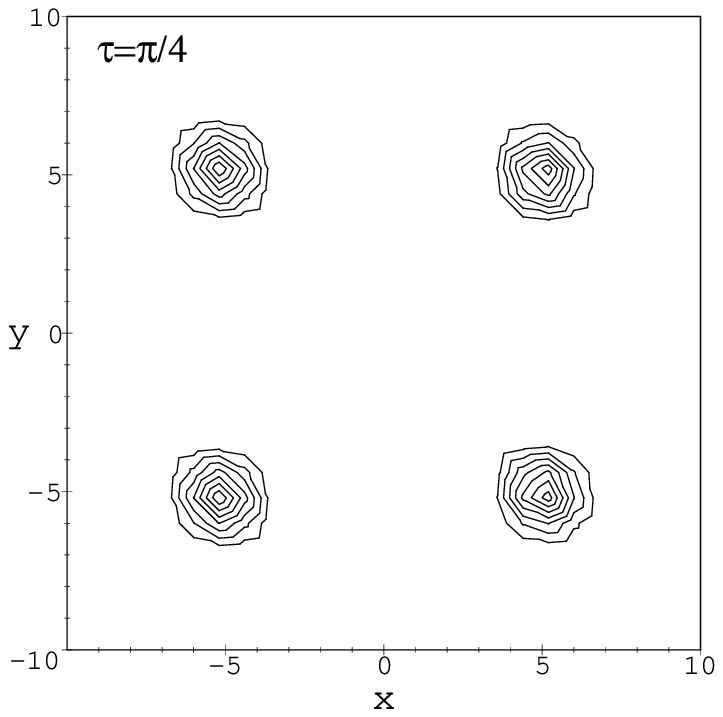}
\epsfxsize=6cm
\epsfbox{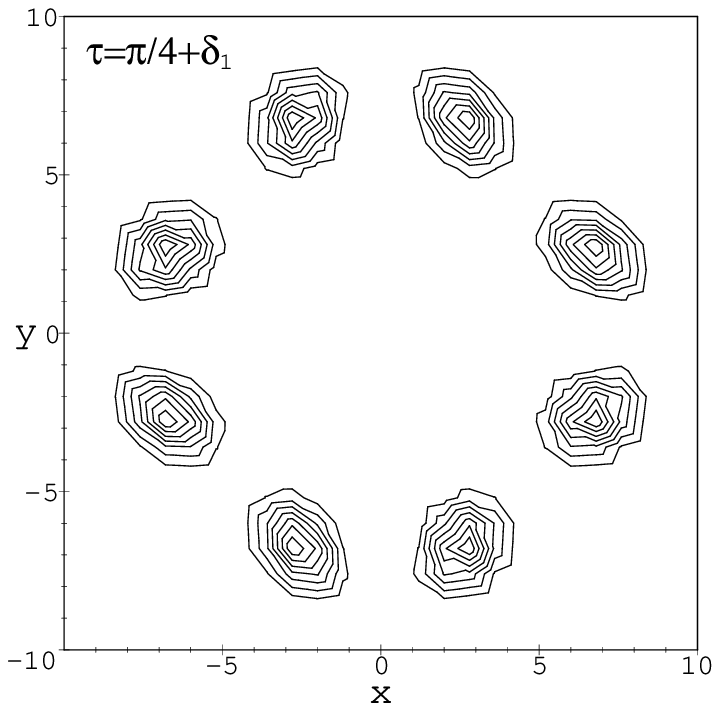}}
\centerline{
\epsfxsize=6cm
\epsfbox{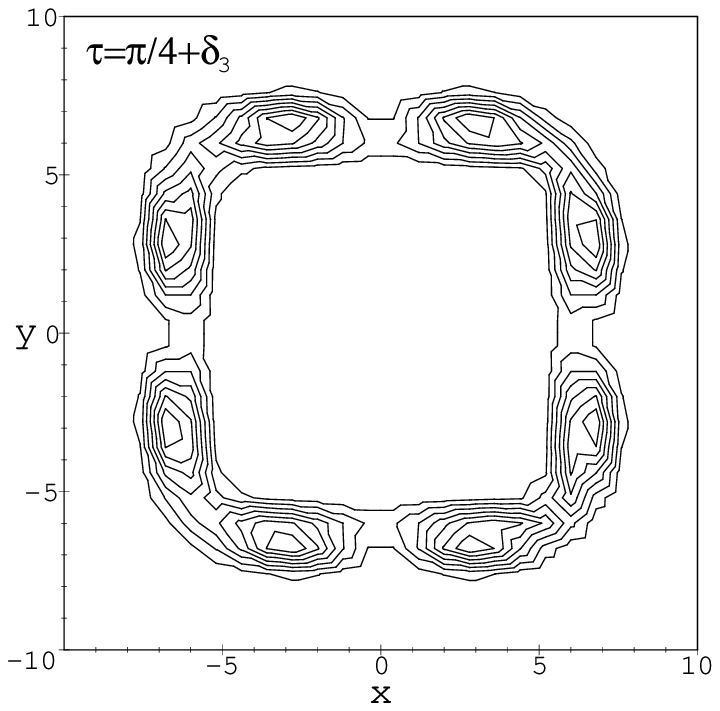}
\epsfxsize=6cm
\epsfbox{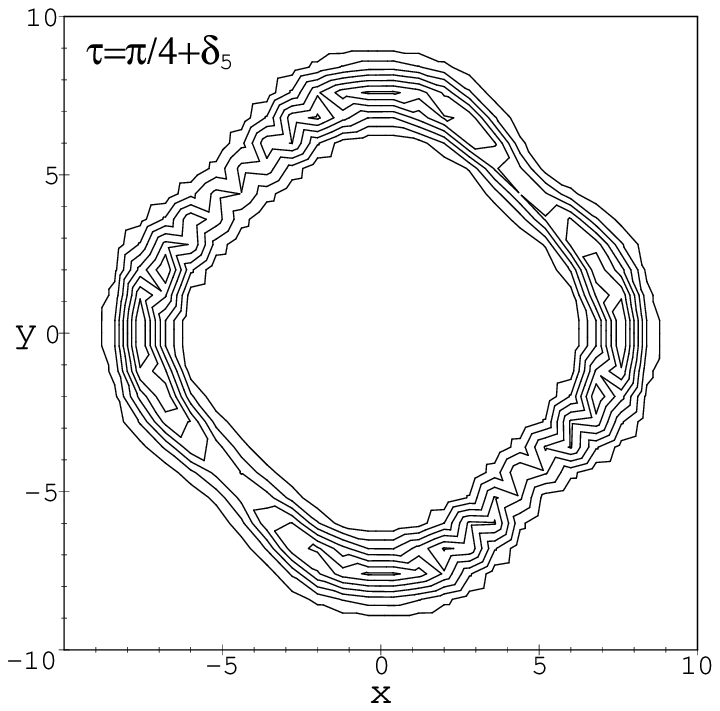}}
\caption{Q-function of the cavity field for $\langle N\rangle=50$
at $\tau=0, \pi/8, \pi/4$ and $\pi$, as well as at $\pi/4+\delta_r$.}
\end{center}
\end{figure}

\begin{figure} \label{inversion}
\begin{center}
\epsfbox{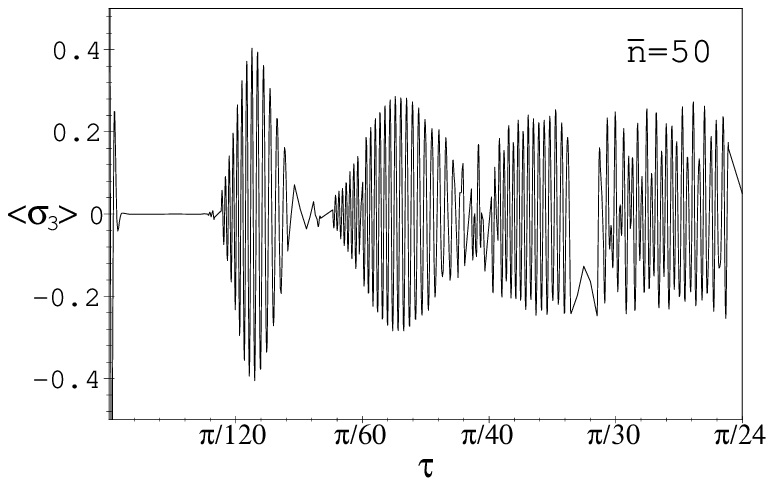}
\caption{Atomic inversion $\langle\sigma_3\rangle$ against $\tau$.}
\end{center}
\end{figure}


\begin{thebibliography}{99} \sc
\bibitem{jay}    Jaynes, E. T., and  Cummings,F. W., 1963, {\it Proc. IEEE, } {\bf 51}, 89.
\bibitem{sho}  Shore, B. W., and  Knight, P. L., 1993, {\it J. Mod. Opt.} {\bf 40}, 1195.
\bibitem{onep}   Gea-Banacloche, J., {\it Phys. Rev. Lett.} {\bf 65} 3385.
\bibitem{buzek} Bu\v{z}ek, V.,  and Hladk\'{y}, B., 1993, {\it J. Mod. Opt. } {\bf 40}, 1309.
\bibitem{fufs}     Fu, H. , Feng, Y., and Solomon, A. I., 2000, {\it J. Phys. A.}, {\bf 33}, 2231.
\bibitem{rabi}     Normally the Rabi frequency is defined as 
                           $g[(n+1)\cdots(n+k)]^{1/2}$. Here, for convenience, we include $g$ in 
                           the definition of the scaled time $\tau=gt$.
\bibitem{Kerr}    Kitagawa, M., and Yamamoto, Y., 1986,
                           {\it Phys. Rev. A},
                          {\bf 34}, 3974;\\
                          Wilson-Gordon, A. D.,  Buzek, V., and Knight, P. L., 1991,  
                          {\it Phys. Rev. A}, {\bf 44}, 7647.
\bibitem{alfred}   Wehrl, A., 1978, {\it Rev. Mod. Phys.}, {\bf 50}, 221.
\bibitem{barnett} Barnett S. M., and Phoenix, S. J. D., 1989, 
                          {\it Phys. Rev. A},  {\bf 40}, 2042.

\bibitem{qfunc}  Hillery, M., O'Connel, R. F., Scully, M. O.,  and Wigner, E. P., 1984,
                           {\it Phys. Rep.}, {\bf 106 }, 121.
\bibitem{newrefs}  Hood, C.J., Chapman, M. S., Lynn, T. W., and Kimble, H. J., 1998,
                              {\it Phys. Rev. Lett.}, {\bf  80}, 4157; 
                              Munstermann P., Fischer, T., Maunz, P., Pinkse, P. W. H., 
                              and Rempe, G.,  1999, {\it Opt. Commun.}, {\bf 159}, 63; 
                              1999, {\it Phys. Rev. Lett.}, {\bf  82}, 3791.
\end{thebibliography}
\end{document}